\def\gsim{\mathrel{\rlap{\lower4pt\hbox{\hskip1pt$\sim$}}
    \raise1pt\hbox{$>$}}}         
\def\lsim{\mathrel{\rlap{\lower4pt\hbox{\hskip1pt$\sim$}}
    \raise1pt\hbox{$<$}}}         

\documentclass{PoS}

\usepackage{url}
\usepackage{cite}
\usepackage{blindtext}

\graphicspath{{plots/}}

\title{Fitting EMC structure functions with intrinsic charm}

\ShortTitle{Fitting EMC structure functions with intrinsic charm}

\author{\speaker{Luca Rottoli}\thanks{On behalf of the NNPDF collaboration: R. D. Ball, V. Bertone, S. Carrazza, L. Del Debbio, S. Forte, P. Groth Merrild, E. R. Nocera, A. Guffanti, N. P. Hartland, Z. Kassabov, J. I. Latorre, J. Rojo, L. Rottoli, M. Ubiali.}\\
        Rudolf Peierls Centre for Theoretical Physics,\\
        1 Keble Road, University of Oxford,\\
        OX1 3NP Oxford, United Kingdom\\
        E-mail: \email{luca.rottoli@physics.ox.ac.uk}\\
        ORCID: {\normalfont \url{http://orcid.org/0000-0001-6967-5127}}
        }

\abstract{ 
A detailed study of the impact of the data collected by the European Muon Collaboration (EMC) on the parton distribution function (PDF) of the charm quark is presented.
The analysis is performed in the NNPDF framework, and the charm PDF is freely parametrized on equal footing as light quark and gluon distributions.
We find that variations in the treatment of EMC data do not modify the charm PDF and do not affect our previous conclusion on the presence of an intrinsic component in the charm PDF.
 }

\FullConference{XXIV International Workshop on Deep-Inelastic Scattering and Related Subjects\\
		11-15 April, 2016\\
		DESY Hamburg, Germany}

\begin{document}

\paragraph{Parton distributions with fitted charm}
A common assumption in modern parton distribution function (PDF) sets~\cite{Harland-Lang:2014zoa,Ball:2014uwa,Dulat:2015mca} is that the charm distribution is perturbatively generated via gluon and light quarks splittings. 
However, this assumption is a possible source of bias for two main reasons. Firstly, the charm PDF may have an `intrinsic', non-perturbative component (see e.g.~\cite{Brodsky:2015fna} and references therein). Secondly, even if the charm PDF is perturbative, its value depends strongly on the value of the charm mass, which is not precisely known.  
To remove this source of bias, the NNPDF collaboration has recently released the NNPDF3IC PDF set where the charm PDF is treated on the same footing as the light quarks and the gluon, i.e. it is fitted from data~\cite{Ball:2016neh}.

In this contribution, we focus on the impact of the EMC data on parton distributions with fitted charm.
We start by reviewing the fit settings and the theory phenomenology used in NNPDF3IC. 
We then analyse the EMC dataset in more detail and we discuss its role in PDF fits with intrinsic charm. 
Finally, we perform a detailed study of the impact of the EMC data on PDFs with fitted charm in the NNPDF framework.

\paragraph{Theory and fitting methodology}
The NNPDF3IC set is largely based on the settings of the NNPDF3.0 global analysis~\cite{Ball:2014uwa}, with some differences regarding theory, fitting methodology, and dataset, which we now briefly discuss.

An accurate treatment of heavy quark mass effects is mandatory in modern PDF sets. 
This is usually achieved resorting to variable flavour number schemes, which ensure a consistent treatment of mass effects close to the quark thresholds, and resummation of large collinear logarithms at large scales. 
The NNPDF family of PDFs uses the FONLL variable flavour number scheme~\cite{Cacciari:1998it,Forte:2010ta} to include heavy quark mass effects.  
Variable flavour number schemes are usually used under the assumption that the heavy quark PDFs are generated perturbatively above the threshold.
However, this assumption is not justified if the heavy quark PDF is freely parameterized, as in the case of fitted charm. 
Therefore, the FONLL scheme has to be extended to include massive charm-initiated contributions~\cite{Ball:2015tna,Ball:2015dpa}.
The new contributions to be added to the FONLL scheme have been initially implemented in the public \texttt{MassiveDISsFunction} code~\cite{MassiveDISsFunction}.
The code has been used as benchmark for the implementation in the public code \texttt{APFEL}~\cite{Bertone:2013vaa}, which has replaced the internal NNPDF \texttt{FKgenerator} code.

Concerning the fitting methodology used in NNPDF3IC, the main difference with respect to NNPDF3.0 is the introduction of an additional PDF in the fitting basis.
Consequently, the seven independent PDF combinations in NNPDF3.0 have been supplemented by the total charm PDF $c^+=c+\bar{c}$ parameterized by a neural network at the initial scale $Q_0$.
We assume that charm and anti-charm PDFs are the same, i.e. $c^-=c-\bar c=0$.
The fits also benefit from minor improvements regarding the positivity constraints, which now include also flavour non-diagonal combinations, and a different treatment of preprocessing exponents.

The experimental data used in the fit is the same as in NNPDF3.0 with two differences. 
Firstly, the combined HERA-I data ~\cite{Aaron:2009aa}
and the HERA-II data from the H1 and ZEUS collaboration~\cite{Aaron:2012qi,Collaboration:2010ry,ZEUS:2012bx,Collaboration:2010xc}
which was included in NNPDF3.0 has been replaced by the HERA legacy data~\cite{Abramowicz:2015mha}.
Secondly, the NNPDF3IC fit includes the EMC charm structure function data~\cite{Aubert:1982tt}.

\paragraph{The EMC data} The EMC charm structure function data has sometimes been adduced as a direct evidence for an intrinsic component in the charm PDF~\cite{Harris:1995jx,Martin:2009iq,Brodsky:1991dj,Brodsky:2015fna} (see also~\cite{Blumlein:2015qcn}).
Nevertheless, previous PDF fits with intrinsic charm reported a poor description of this data~\cite{Jimenez-Delgado:2014zga}.    
It is now known that data from the EMC experiment suffered from systematic uncertainties which had not been fully taken into account at the time of the measurement. 
Consequently, some argue that the reliability of the EMC data is questionable. 
The main reason of concern is the tension~\cite{Feltesse:2012yaa} between the inclusive EMC structure function data~\cite{Aubert:1985fx,Aubert:1987da} and the data from the BCDMS experiment~\cite{bcdms1,bcdms2}.  
Furthermore, the charm structure function data from the EMC collaboration was extracted using an inclusive branching ratios (BR) of D mesons into muons equal to $8.2 \%$, which differs from the last measurement of LHCb~\cite{Gauld:2015yia,Agashe:2014kda} as well as the PDG average~\cite{Agashe:2014kda}.

The NNPDF3IC set includes charm structure function data from the EMC collaboration extracted from di-muon and tri-muon events produced by the interaction of muons with $E_\mu=250$ GeV in an iron target.  
This data should be unaffected from the underestimated systematic uncertainties, which instead are related to the drift chambers which were not used for this measurement.\footnote{We acknowledge discussions with M. Arneodo for detailed information on the EMC charm structure function dataset.
} 
In order to quantify the impact of the 250 GeV data several variations have been performed in~\cite{Ball:2016neh}. 
In particular, the results obtained with and without EMC data have been compared, as well as the results obtained by rescaling the EMC data by taking into account the updated branching ratios. 
It was concluded that whereas discarding completely the EMC charm data has a significant effect in the charm PDF at medium and small values of $x$, different treatments of the dataset do not alter substantially the results obtained with the original dataset.

\paragraph{The impact of the EMC data}
In this contribution, we show some of the results of the study performed in~\cite{Ball:2016neh} in greater detail, and we investigate the impact of other EMC datasets.
More specifically, we include the EMC charm structure function data extracted from di-muon and tri-muon events with 200 GeV muons of ref.~\cite{Arneodo:1986nj}.
To our knowledge, this data has never been considered in the context of intrinsic charm studies.
The 200 GeV dataset, despite the lower statistics, provides additional constraints and therefore should be used alongside the 250 GeV dataset.
Finally, we assess the impact of the inclusive EMC structure function data of ref.~\cite{Aubert:1985fx,Aubert:1987da}. 

\begin{figure}[t]
  \includegraphics[width=0.495\textwidth]{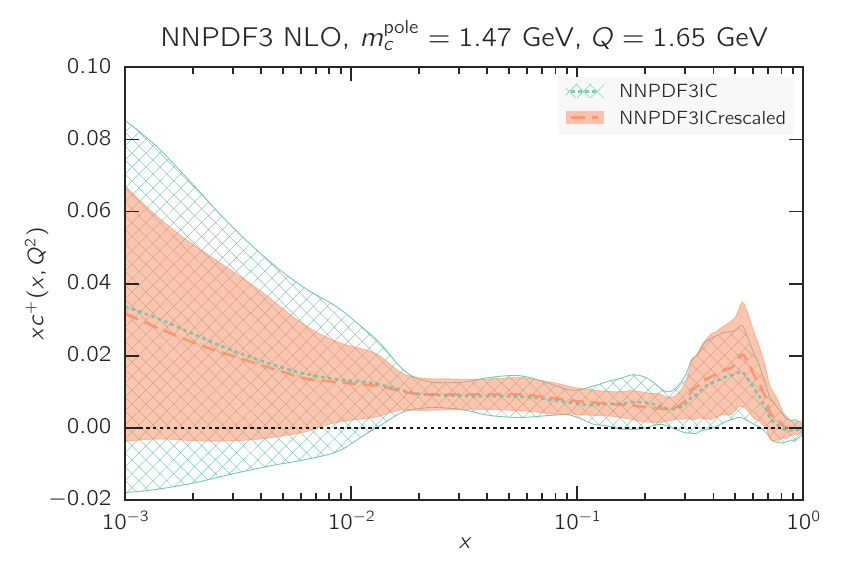}
  \includegraphics[width=0.495\textwidth]{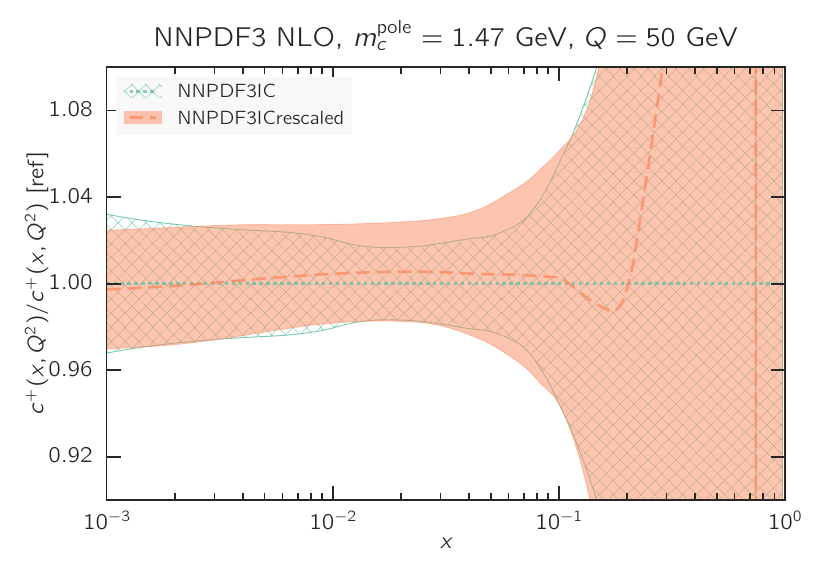}
  \caption{Comparison of the NNPDF3IC PDF set to the PDF set obtained with up-to-date branching ratio, for a charm pole mass of $m_c^{\rm{{pole}}} = 1.47$ GeV. Left panel: comparison of the charm PDF at $Q=1.65$ GeV. Right panel: comparison of the charm PDF at $Q=50$ GeV. The comparison at $Q=50$ GeV is shown as a ratio.}
  \label{fig:1}
\end{figure}

Let us start to discuss the impact of using the up-to-date branching ratio of D mesons into muons in a PDF set with fitted charm. 
To quantify this impact we rescaled the EMC data by a factor of 0.82 and we added an additional 15$\%$ systematic uncertainty, which we assumed as fully correlated.
The effect on the charm PDF is shown in Fig.~\ref{fig:1} at two different scales. 
The rescaling has a negligible impact on the central value of the charm PDF. 
The PDF uncertainty at $Q=1.65$ GeV is somewhat reduced, although compatibly with statistical fluctuation.
We observe that the large-$x$ structure of the charm PDF at $Q=1.65$ GeV is the same regardless of the rescaling.
The impact of the rescaling on the other PDFs is negligible and the overall quality of the fit is unchanged. 
Henceforth, we will use the fit obtained with the up-to-date branching ratio as a baseline.

\begin{figure}[t]
  \includegraphics[width=0.495\textwidth]{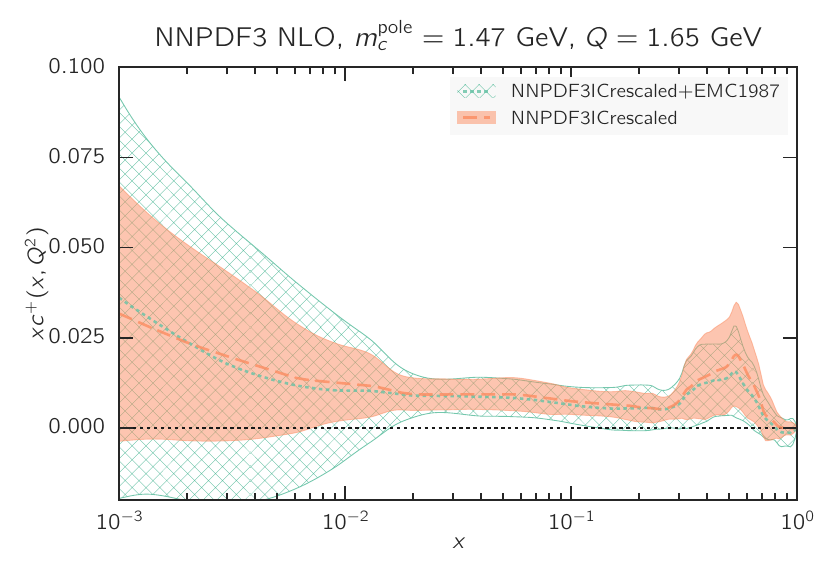}
  \includegraphics[width=0.495\textwidth]{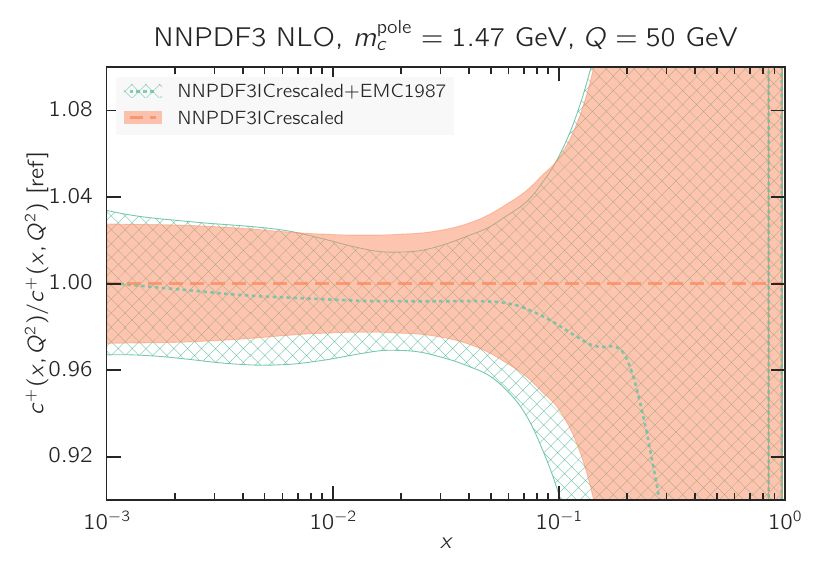}
  \caption{Same as Fig.~\protect\ref{fig:1}, now comparing the baseline fit with a fit including the 200 GeV EMC charm structure function data.}
  \label{fig:2}
\end{figure}

We can now compare our baseline fit with a PDF set extracted using all the charm structure function data collected at the EMC experiment. 
The 200 GeV data has been extracted using the same BR used in the 250 GeV dataset, therefore we applied the rescaling procedure to this dataset as well. 
The additional 15$\%$ systematic uncertainty has been assumed as fully correlated between the two datasets.
The results are shown in Fig.~\ref{fig:2}. 
We observe that the central value of the charm PDF is substantially unchanged.
The PDF uncertainty of the charm PDF is slightly larger if the 200 GeV dataset is included, especially at smaller values of $x$, again compatibly within statistical fluctuations.
The PDF at large $x$ is essentially unaltered by the inclusion of the new data.

We can finally assess the impact of the inclusive EMC data. 
To that purpose, we have performed a new fit which includes both the 250 GeV and 200 GeV charm structure function data and the inclusive EMC data. 
As expected, the quality of the fit partially deteriorates due to the tension between this data and the other data included in the global fit.
The description of the EMC inclusive dataset is rather unsatisfactory, with a $\chi^2$ equal to 4.8 per datapoint.
However, the impact of the data on the fit is moderate, and the $\chi^2$ of the other experiments fluctuates but does not worsen.
We have performed a variation of this last fit by imposing an additional cut on the inclusive dataset and discarding all the experimental points with $x<0.1$.
The systematic uncertainties which affected the drift chambers should in fact have a more pronounced effect on the small-$x$ datapoints.
Indeed, the description of the inclusive EMC data with this additional cut improves, although the description of the data is not completely satisfactory as the $\chi^2/\textrm{d.o.f.}$ is equal to $2.8$.

\begin{figure}[t]
  \includegraphics[width=0.495\textwidth]{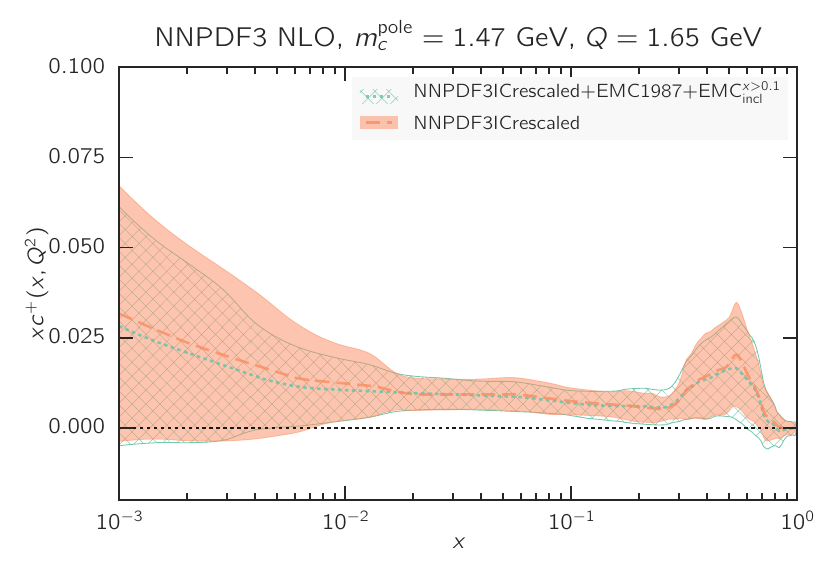}
  \includegraphics[width=0.495\textwidth]{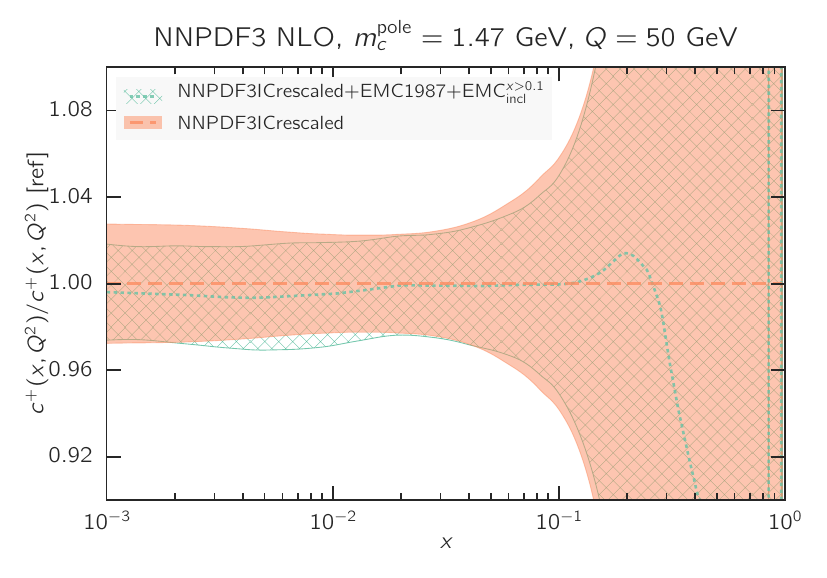}
  \caption{Same as Fig.~\protect\ref{fig:1}, now comparing the baseline fit with a fit including the 200 GeV EMC charm structure function data and the inclusive EMC data.}
  \label{fig:3}
\end{figure}

We show in Fig.~\ref{fig:3} the impact of the inclusive EMC data on the charm PDF at 1.65 GeV and at 50 GeV. 
The impact of the inclusive EMC dataset is marginal: the central value of the charm PDF is not affected by the addition of the EMC inclusive dataset, and we observe only small differences in the PDF uncertainty in the small-$x$ region with regard to the baseline fit. 
We observe a partial reduction in the PDF uncertainty at small-$x$ in respect of the fit with the 200 GeV EMC charm structure function data (see Fig.~\ref{fig:2}).
The results are comparable within statistical uncertainties.

\paragraph{Discussion and oulook}To conclude, we have studied in detail the impact of the EMC data on parton distributions with fitted charm.
We have included all the charm structure function data collected by the EMC collaboration.
We have shown that different treatments of the EMC charm structure functions data do not substantially alter the fitted charm PDF.
Finally, we have inspected the impact of the inclusive EMC structure function data. 
Whereas this data cannot be satisfactorily described, its impact on the fit is rather moderate, especially if only the large-$x$ data is included in the fit.
Furthermore, the impact of this data on the charm PDF is small.
We found that the large-$x$ structure in the charm PDF is particularly stable upon all the variations we have performed.
This structure is absent in case the charm PDF is assumed to be generated perturbatively, thereby suggesting the presence of an intrinsic component peaked at large $x$ in the charm PDF.

\paragraph{Acknowledgements} I would like to thank Amanda Cooper-Sarkar, Pavel Nadolsky and Robert Thorne for challenging discussions regarding the EMC data. 
I also acknowledge countless discussions with the members of the `extended' NNPDF collaboration, in particular Valerio Bertone, Marco Bonvini, Nathan Hartland and Juan Rojo.
Finally, I am pleased to thank the organizers of the DIS 2016 workshop for the opportunity to spend a lovely week in DESY Hamburg.
In this time of trouble, I feel necessary to thank all the EU taxpayers as this work has been supported by an European Research Council Starting Grant "PDF4BSM". %

\bibliography{rottoli_DIS}

\end{document}